\documentclass[%
 aip,
 jmp,%
 amsmath,amssymb,
preprint,%
]{revtex4-1}

\usepackage{graphicx}
\usepackage{dcolumn}
\usepackage{bm}

\begin{document}


\title[Unobservable Potentials to Explain a Quantum Eraser and a Delayed-Choice Experiment]{Unobservable Potentials to Explain a Quantum Eraser and a Delayed-Choice Experiment}

\author{Masahito Morimoto}
 \altaffiliation{Telecommunications \& Energy Laboratories of Furukawa Electric Co., Ltd. \\ 6, Yawata-Kaigandori, Ichihara, Chiba, 290-8555, Japan \\}
\email{masahito.s.morimoto@furukawaelectric.com}



\begin{abstract}
We present a new explanation for a quantum eraser. Mathematical description of the traditional explanation needs quantum-superposition states.
However, the phenomenon can be explained without quantum-superposition states by introducing unobservable potentials which can be identified as an indefinite metric vector. In addition, a delayed choice experiment can also be explained by the interference between the photons and unobservable potentials, which seems like an unreal long-range correlation beyond the causality.%
%
\end{abstract}

\maketitle
\section{Introduction}
Quantum theory has paradoxes related to the reduction of the wave packet typified by ``Schr\"odinger's cat'' and ``Einstein, Podolsky and Rosen (EPR)'' \cite{Schr-cat,EPR}. In order to interpret the quantum theory without paradoxes, de Broglie and Bohm had proposed so called ``hidden variables'' theory \cite{Bohm1,Bohm2}. Although ``hidden variables'' has been rejected at that time\cite{Bell}, the theory has been improved in a way that is consistent with relativity and ontology \cite{Bohmian1,Bohmian2,Bohmian3,Bohmian4,Hiley}. However, the improvement has not been completed so far.

Several experiments have demonstrated that Bell's inequalities are always violated confirming the quantum mechanics theory on the non-locality of the photon and demonstrating the absence of ``hidden variables'' for the local representation \cite{aspect1981experimental,aspect1982experimental,aspect1982experimental-2}.

However, the author has reported the alternative interpretation for quantum theory utilizing quantum field formalism with unobservable potentials \cite{Morimoto} that can be identified as ``hidden variables'' similar to Aharonov-Bohm effect \cite{0370-1301-62-1-303,ABeffe} and rigorous mathematical treatment using tensor form in keeping with the local representation, i. e., consistent with relativity. The interpretation can omit the quantum paradoxes and be applicable to elimination of infinite zero-point energy, spontaneous symmetry breaking, mass acquire mechanism, non-Abelian gauge fields and neutrino oscillation, which can lead to the comprehensive theory.

The alternative interpretation gives completely the same calculation results using the traditional quantum-superposition states because the mathematical tools involved in the calculations, such as routine state vectors, operators, and inner products, are identical to those used in traditional quantum theory. The difference between the alternative and traditional treatment is the introduction of indefinite metric as a physical reality that contradicts ``probabilistic interpretation''. In the alternate interpretation, the inner product of the states which has been recognized as so called ``probability amplitudes'' is unrelated to the probability but related to an amplitude of interferences. Hence the ``interference amplitudes'' is preferable to ``probability amplitudes'', though we will use the word ``probability amplitudes'' in this paper according to traditional way. Although the calculation method of the alternative interpretation in this paper using covariant quantization might be slightly more complicated than the traditional one without covariant quantization, the method is a straightforward approach, and the result is an inevitable conclusion by the rigorous derivation. 

In one example, the linear equations, e. g., Maxwell and Schr\"odinger equations, allow a superposition of any eigenfunctions as a different solution. An eigenfunction can represent an eigenstate after quantization, which describes non-divisible eigenstates such as single photon and electron. Although the superposition states are allowed by the linearity of the equations, the non-divisible eigenstate should not be divided after quantization, i. e., the coefficient so-called ``probability amplitude'' of the non-divisible eigenstate must be integer. In other words, the eigenstates are just mode eigenfunctions derived from the geometry (boundary condition of the equations), and the superposition states composed of broken eigenstates should not be configured for an initial condition after quantization. Therefore, the superposition of the eigenstates whose coefficients are not integer has to be recognized as statistical treatment in mixed states for the case that a lot of particles exist, e. g., the normalization of the coefficients is obviously the statistical treatment which allows probabilistic interpretation.
 
However, in order to justify the phenomena looking like the quantum superposition states in the
case that few particles exist such as single particle, we need some infinitely divisible (i.e., arbitrary coefficient) continuous body regardless of the quantization. The author find that the unobservable (scalar) potentials must be just the thing which acts as a substitute for the superposition as an inevitable result from the rigorous covariant quantization without any artificial treatment. The result is not a matter of interpretation or the authors claims but just findings.

Here we introduce an example of the findings as reported in Ref.\cite{Morimoto}, and two-path single photon and electron interference can be calculated without quantum-superposition state by introducing a substantial (localized) photon or electron and the unobservable (scalar) potentials, which are expressed as following Maxwell equations. 
\begin{eqnarray}
\left( \Delta -\frac{1}{c^{2}} \frac{\partial ^{2}}{\partial t^{2}} \right) {\bf A} - \nabla \left( \nabla \cdot {\bf A} + \frac{1}{c^{2}} \frac{\partial \phi}{\partial t} \right)  & = & - \mu_{0} {\bf i} \nonumber \\
\left( \Delta -\frac{1}{c^{2}} \frac{\partial ^{2}}{\partial t^{2}} \right) \phi + \frac{\partial}{\partial t} \left( \nabla \cdot {\bf A} + \frac{1}{c^{2}} \frac{\partial \phi}{\partial t} \right) & = & - \frac{\rho}{\varepsilon_{0}}
\label{eq:Maxl}
\end{eqnarray} 
When the scalar potential of Eq. (\ref{eq:Maxl}) is quantized, the photon annihilation operator $ \hat{A}_{0}' $ expressing the unobservable (scalar) potential can be expressed as follows.
\begin{eqnarray}
\hat{A}_{0}' = \frac{1}{2} \gamma e^{i \theta/2} \hat{A}_{1} - \frac{1}{2} \gamma e^{- i \theta/2} \hat{A}_{1} & , \ \ \ & \hat{A}_{0}'^{\dagger} = \frac{1}{2} \gamma e^{- i \theta/2} \hat{A}_{1}^{\dagger} - \frac{1}{2} \gamma e^{ i \theta/2} \hat{A}_{1}^{\dagger}
\label{eq:a2}
\end{eqnarray}
where $ \gamma^{2} = - 1 $ ( i. e., $ \gamma $ corresponds to the square root of the determinant of Minkowski metric tensor $ \sqrt{ | g_{\mu \nu} | } \equiv \sqrt{g} \equiv \sqrt{-1} = \gamma $) which stands for requirement of indefinite metric, $ \hat{A}_{1} $ is the photon annihilation operator obtained from quantization of the vector potentials in Eq. (\ref{eq:Maxl}); $ \theta $ is a phase difference derived from a geometry. By using tensor form (covariant quantization), we can explicitly identify these operators $ \hat{A}_{0}' $ as the scalar potential and $ \hat{A}_{1} $ as the vector potentials. This description is spontaneously obtained as described later.

The above $ \hat{A}_{0}' $ is quite similar to the expression of $\tilde{{\bf \Xi}} $ reported by Meis to investigate quantum vacuum state as follows \cite{Meis}.
\begin{equation}
\tilde{{\bf \Xi}}_{0_{k \lambda}} = \xi a_{k \lambda} \hat{\epsilon}_{k \lambda} e^{i \varphi} + \xi^{*} a^{\dagger}_{k \lambda} \hat{\epsilon}^{*}_{k \lambda} e^{-i \varphi}
\label{eq:meis}
\end{equation}
where $ k $, $ \lambda $, $ \epsilon $, $ \xi $ and $ \varphi $ stand for $ k $ mode, $ \lambda $ polarization, a complex unit vector of polarization, a constant and a phase parameter, respectively.

If we identify $ \xi $ and $ \xi^{*} $ as $ \frac{1}{2} \gamma $ and $ -\frac{1}{2} \gamma $ and  introduce polarization vectors as described later in Eq. (\ref{eq:90and0do}), then Eq. (\ref{eq:a2}) corresponds to Eq. (\ref{eq:meis}).

When state vector $ | \zeta \rangle $, which represents the unobservable (scalar) potentials, is introduced in Schr\"odinger picture as follows, the vector can be identified as indefinite metric vector.
\begin{equation}
| \zeta \rangle \equiv \left( \frac{1}{2} \gamma e^{i \theta/2} - \frac{1}{2} \gamma e^{- i \theta/2} \right) | 1 \rangle
\label{eq:zeta}
\end{equation}
where $ | 1 \rangle $ represents a photon state. Therefore when there is no phase difference, the expectation value of arbitrary physical quantity $ \hat{A} $ and probability (or more like ``interference'') amplitude of $ | \zeta \rangle $ are zeros ($ \langle \zeta | \hat{A} | \zeta \rangle = 0 $ , $ \langle \zeta | \zeta \rangle = 0 $), which means the unobservable potentials can not be observed alone in the literature. More detail treatment of these operators and vectors have been discussed in Ref.\cite{Morimoto}.

Aharonov and Bohm have pointed out that the unobservable potentials can cause electron wave interferences \cite{ABeffe}, and we should realize that all of physical interactions are regulated by gauge fields (gauge principle. the potentials are also gauge fields.), which can not be observed alone \cite{Yang1,Yang2,Weinberg,Utiyama}.

In this paper, we show that the existence of the unobservable potentials can explain not only the interferences but also the quantum eraser and delayed choice experiment. In addition, we also show that the interference between photons and the unobservable potentials violates Bell's inequalities in keeping with the local representation, which is consistent with relativity. This fact is the most important novel aspect of this paper that the violation of Bell's inequalities can not justify the non-locality of quantum theory and the absence of ``hidden variables'' because the unobservable potentials which propagate through space at the speed of light, i. e., ``local action'' or ``locality'', can act as ``hidden variables''.

\section{Traditional explanation for quantum eraser}

\begin{figure}[t]
\centering
\includegraphics[width=18pc]{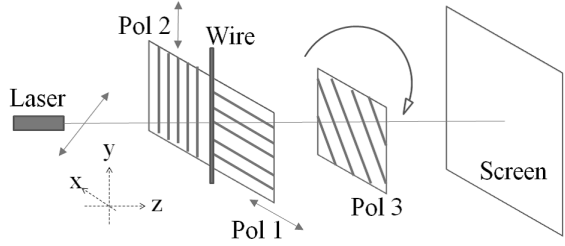}
\caption{Typical setup for the Quantum Eraser. Pol1 and Pol2 are fixed linear polarizers with polarizing axes perpendicular (x and y). Pol3 is a revolvable linear polarizer.} 
\label{fig:one}
\end{figure}

Figure \ref{fig:one} shows a typical setup for the quantum eraser \cite{Doitysqe}. 
When there are no polarizers, an interference pattern composed of dark and bright fringes can be observed on the screen because light passing on the left of the wire is combining, or ``interfering,'' with light passing on the right-hand side. In other words, we have no information about which path each photon went.

When polarizers 1 and 2, which are called ``which-path markers'', are positioned right behind the wire as shown in Figure \ref{fig:one}, the launched light polarized in 45$^\circ $ direction from the Laser is polarized in perpendicular (x-polarized and y-polarized) by these polarizers. Then the interference pattern on the screen is erased because ``which-path makers'' have made available the information about which path each photon went.

When polarizer 3 is inserted in front of the screen with the polarization angle +45$^\circ $ or -45$^\circ $ in addition to ``which-path makers'', the interference pattern reappears because polarizer 3 has made the information of ``which-path makers'' unusable.

We can produce a mathematical description of the erasure and reappearance of the interference pattern as follows.
The x- and y-polarized photon passing through polarizer 1 and 2 can be expressed by the quantum-superposition state as follows.  
\begin{equation}
| x \rangle = \frac{1}{\sqrt{2}} | + \rangle + \frac{1}{\sqrt{2}} | - \rangle
\label{eq:90do}
\end{equation}
and
\begin{equation}
| y \rangle = \frac{1}{\sqrt{2}} | + \rangle - \frac{1}{\sqrt{2}} | - \rangle
\label{eq:0do}
\end{equation}

where ``+'' and ``-'' represent polarizations +45$^\circ $ and -45$^\circ$ with respect to $ x $.

The photons pass through polarizers 1 and 2 are polarized at right angles to each other as seen in the left-hand side of Eqs. (\ref{eq:90do}) and (\ref{eq:0do}), which prevent the interference pattern. In other words, ``which-path makers'' have made available the information about which path each photon went. Although there are same polarized states in the right-hand side of Eq. (\ref{eq:90do}) and (\ref{eq:0do}), the interference patterns consisting of bright and dark fringes made by +45$^\circ $ and -45$^\circ $ polarized states are reverted images and annihilate each other. Therefore, sum total of the images has no interference pattern.

When polarizer 3 is inserted with the polarization angle +45$^\circ $ or -45$^\circ $, only $ | + \rangle $ or $ | - \rangle $ can pass through polarizer 3. Then the interference pattern made by either $ | + \rangle $ or $ | - \rangle $ of both Eqs. (\ref{eq:90do}) and (\ref{eq:0do}) reappears, which means that we can not identify which-path the photons had passed through, i.e., polarizer 3 has made the information of ``which-path makers'' unusable.

\section{New explanation for quantum eraser}

The mathematical description of the photon states passing through polarizers 1 and 2 used in the traditional explanation requires the quantum-superposition states in Eqs. (\ref{eq:90do}) and (\ref{eq:0do}), respectively.

If Maxwell equations are deemed to be classical wave equations whose electro-magnetic fields obey the superposition principle, then the description is valid. However, applying the superposition principle to particle image, e. g., inseparable single photon, leads to quantum paradoxes such as the reduction of the wave packet. These paradoxes are great problems not only with the traditional explanations but also for true nature of physics. 

Although tensor form (covariant quantization) is a rigorous treatment as we will describe later,
here we conveniently take advantage of the unobservable potentials that can eternally populate the whole space as waves independent of existence of the substantial photons. Therefore, we can replace the photon state $ | x \rangle $ with $ | x \rangle + | \zeta \rangle $, where $ | \zeta \rangle $ is a state representing the unobservable potentials whose probability (or more like ``interference'') amplitudes $ \langle \zeta | \zeta \rangle = 0 $ in initial states as described in Eq. (\ref{eq:zeta}) (when there is no difference in phase and polarization angle as described below.). The unobservable potentials can be polarized by the polarizers because these potentials obey Maxwell equations and populate the whole space-time. Therefore, we should introduce the polarization terms with unobservable potentials.

Then the following states, which are identified as Eq. (\ref{eq:zeta}) introducing polarization terms similar to Eq. (\ref{eq:meis}), can generate the same interference as the quantum-superposition states in Eqs. (\ref{eq:90do}) and (\ref{eq:0do}).
\begin{eqnarray}
| x \rangle + | \zeta_{\phi, x} \rangle & = & | x \rangle + \frac{1}{2} \gamma e^ {i \phi} e^{i \theta/2}  | x \rangle - \frac{1}{2} \gamma e^{- i \phi} e^{- i \theta/2}  | x \rangle  \nonumber \\
| y \rangle + | \zeta_{\phi + \frac{1}{2} \pi, y} \rangle & = & | y \rangle + \frac{1}{2} \gamma e^ {i (\phi + \frac{1}{2}\pi )} e^{- i \theta/2}  | y \rangle - \frac{1}{2} \gamma e^{- i  (\phi + \frac{1}{2}\pi )} e^{i \theta/2}  |  y \rangle  
\label{eq:90and0do}
\end{eqnarray}
where $ \gamma^2 = -1 $, $ \phi $ and $ \theta $ are the indefinite metric, the polarization angle of polarizer 3 measured from x-axis and phase difference between left and right paths, respectively.

Therefore, when we observe only $ | x \rangle $ with polarizer 3, i. e., $ \theta = 0 $, the intensity of the interference $ \langle I \rangle $ can be calculated as follows.
\begin{eqnarray}
\langle I \rangle \propto  \left( \langle x | + \langle \zeta_{\phi, x} | \right) \left( | x \rangle + | \zeta_{\phi, x} \rangle \right) & = & \langle x | x \rangle - \frac{1}{2} \langle x | x \rangle + \frac{1}{2} \langle x | x \rangle \cos\left( 2 \phi + \theta \right) \nonumber \\
& = & \frac{1}{2} + \frac{1}{2} \cos\left( 2 \phi + \theta \right)  = \frac{1}{2} + \frac{1}{2} \cos\left( 2 \phi \right)
\label{eq:intensity-x}
\end{eqnarray}
Hence the output intensity by rotation angle of polarizer 3 is reproduced correctly.

When we observe $ | x \rangle $ and $ | y \rangle $ with polarizer 3, the intensity is obtained as follows.
\begin{equation}
\langle I \rangle \propto \left( \langle x | + \langle \zeta_{\phi, x} | + \langle y | + \langle \zeta_{\phi + \frac{1}{2}\pi, y} | \right) \left( | x \rangle + | \zeta_{\phi, x} \rangle + | y \rangle + | \zeta_{\phi + \frac{1}{2}\pi, y} \rangle \right)
\label{eq:intensity-xy}
\end{equation}
Because $ \langle x | y \rangle = \langle y | x \rangle = 0 $,
\begin{equation}
\langle I \rangle \propto \left( \langle x | + \langle \zeta_{\phi, x} | \right) \left( | x \rangle + | \zeta_{\phi, x} \rangle \right) + \left( \langle y | + \langle \zeta_{\phi + \frac{1}{2}\pi, y} | \right) \left( | y \rangle + | \zeta_{\phi + \frac{1}{2}\pi, y} \rangle \right) 
\end{equation}
By using Eq. (\ref{eq:intensity-x}), we can obtain the following result.
\begin{equation}
\langle I \rangle \propto  \frac{1}{2} + \frac{1}{2} \cos\left( 2 \phi + \theta \right) + \frac{1}{2} + \frac{1}{2} \cos\left( 2 \phi + \pi - \theta \right) = 1 +  \frac{1}{2} \cos\left( 2 \phi + \theta \right) - \frac{1}{2} \cos\left( 2 \phi - \theta \right)
\label{eq:intensity-xy-comp}
\end{equation}
When $ \phi = \pm \pi, \ \pm \frac{1}{2} \pi $, $ \langle I \rangle \propto 1 $ and $ \phi = \pm \frac{1}{4} \pi $, then $ \langle I \rangle \propto 1 \pm \sin \theta $, which reproduces the interference correctly.

In this new explanation, the polarization of substantial photons is fixed, and the photons can not pass through the polarizer which has a different polarization angle. However, the unobservable potentials create the same interference as the superposition state of $ | + \rangle $ and $ | - \rangle $ as described above. In the case of x-polarized single photon, the interference can be calculated by Eq. (\ref{eq:90and0do}) replacing $ | y \rangle $ with $ | 0 \rangle $. Then $ \langle I \rangle \propto  1 +  \frac{1}{2} \cos\left( 2 \phi + \theta \right) - \frac{1}{2} \cos\left( 2 \phi - \theta \right)$ is obtained.  Note that when we calculate the single photon interference by using photon number operator $ {\bf n}_{1} = \hat{A}_{1}^{\dagger} \hat{A}_{1} $, we can obtain exact expression $ \langle I \rangle \propto  \frac{1}{2} +  \frac{1}{2} \cos\left( 2 \phi + \theta \right) $ because $ \langle 0 | 0 \rangle = 1 \neq \langle 0 | {\bf n}_{1} | 0 \rangle = 0 $, where $ \hat{A}_{1} $ is the photon annihilation operator obtained from the vector potentials in Eq. (\ref{eq:Maxl}) \cite{Morimoto}.

The above calculations are based on Schr\"odinger picture. We can obtain the same results based on Heisenberg picture. In Heisenberg picture, the photon number operator should be replaced by $ {\bf n} = ( \hat{A}_{1}^{\dagger} + \hat{A}_{p}^{\dagger} )( \hat{A}_{1} + \hat{A}_{p} ) $ \cite{Morimoto}, where $ \hat{A}_{1} $ and $ \hat{A}_{p} \ ( p : {\rm polarization} = x , \ y , \cdots , {\rm etc.} ) $ are the photon annihilation operators obtained from the vector and scalar potentials in Eq. (\ref{eq:Maxl}), respectively, which represents the substantial photons and modified operator introducing the polarization terms in Eq. (\ref{eq:a2}), i. e., the polarized unobservable potentials, as follows.
\begin{eqnarray}
\hat{A}_{x} = \frac{1}{2} \gamma e^ {i \phi} e^{i \theta/2} \hat{A}_{1} - \frac{1}{2} \gamma e^ {- i \phi} e^{- i \theta/2} \hat{A}_{1} & , \ \ \ & \hat{A}_{x}^{\dagger} = \frac{1}{2} \gamma e^ {- i \phi} e^{- i \theta/2} \hat{A}_{1}^{\dagger} - \frac{1}{2} \gamma e^ { i \phi} e^{ i \theta/2} \hat{A}_{1}^{\dagger}
\label{eq:a2-phi}
\end{eqnarray}
We can calculate Eq. (\ref{eq:intensity-x}) in Heisenberg picture as follows.
\begin{eqnarray}
\langle I \rangle & = & \langle n | ( \hat{A}_{1}^{\dagger} + \hat{A}_{x}^{\dagger} )( \hat{A}_{1} + \hat{A}_{x} ) | n \rangle \nonumber \\
& = & \langle n | {\bf n}_{1} | n \rangle + \langle n | \hat{A}_{x}^{\dagger} \hat{A}_{x} | n \rangle \propto 1 - \frac{1}{2} + \frac{1}{2} \cos\left( 2 \phi + \theta \right)  = \frac{1}{2} + \frac{1}{2} \cos\left( 2 \phi \right)
\label{eq:intensity-x-Heisen}
\end{eqnarray}
Note that the x-polarized photon annihilation operator should be represented by $ \hat{A}_{1} + \hat{A}_{x} $ instead of $ \hat{A}_{1} $ in Heisenberg picture \cite{Morimoto}. When there are x- and y-polarized photons, the operator should be represented by $ (\hat{A}_{1} + \hat{A}_{x}) + (\hat{A}_{2} + \hat{A}_{y}) $, where $ \hat{A}_{2} $ is a photon annihilation operator obtained from the quantization of y-polarized vector potential, and $ \hat{A}_{y} $ can be obtained by replacing $ \phi $ with $ \phi + \frac{1}{2} \pi $ and $ \hat{A}_{1} , \hat{A}_{1}^{\dagger} $ with $ \hat{A}_{2} ,\hat{A}_{2}^{\dagger} $ in Eq. (\ref{eq:a2-phi}). 
Then we can calculate Eq. (\ref{eq:intensity-xy}) in Heisenberg picture as follows.
\begin{eqnarray}
\langle I \rangle & = & \langle n | ( \hat{A}_{1}^{\dagger} + \hat{A}_{x}^{\dagger} + \hat{A}_{2}^{\dagger} + \hat{A}_{y}^{\dagger} )( \hat{A}_{1} + \hat{A}_{x} + \hat{A}_{2} + \hat{A}_{y} ) | n \rangle \nonumber \\
& = & \langle n | {\bf n}_{1} | n \rangle + \langle n | \hat{A}_{x}^{\dagger} \hat{A}_{x} | n \rangle + \langle n | {\bf n}_{2} | n \rangle + \langle n | \hat{A}_{y}^{\dagger} \hat{A}_{y} | n \rangle \nonumber \\
& \propto & 1 +  \frac{1}{2} \cos\left( 2 \phi + \theta \right) - \frac{1}{2} \cos\left( 2 \phi - \theta \right)
\label{eq:intensity-xy-Heisen}
\end{eqnarray}
where we identify $ \langle n | {\bf n}_{1} | n \rangle \equiv \langle n | \hat{A}_{1}^{\dagger} \hat{A}_{1} | n \rangle = \langle n | {\bf n}_{2} | n \rangle \equiv \langle n | \hat{A}_{2}^{\dagger} \hat{A}_{2} | n \rangle = n $ assuming that there are the same number ($n$) of x- and y-polarized photons. Under the assumption $ | n \rangle \equiv | n \rangle_{x} + | n \rangle_{y} $ where $ | n \rangle_{x}, | n \rangle_{y }$ are the x- and y-polarized $ n $ photon states, respectively, we can calculate $ \hat{A}_{1} | n \rangle =  \hat{A}_{1} | n \rangle_{x} + \hat{A}_{1} | n \rangle_{y} = \sqrt{n} | n -1 \rangle_{x} $ and $ \hat{A}_{2} | n \rangle =  \hat{A}_{2} | n \rangle_{x} + \hat{A}_{2} | n \rangle_{y} = \sqrt{n} | n -1 \rangle_{y} $. In addition, $ \langle n | \hat{A}_{1}^{\dagger} \hat{A}_{2} | n \rangle = \langle n | \hat{A}_{2}^{\dagger} \hat{A}_{1} | n \rangle = 0 $ is calculated.

The new explanation can describe that $ \hat{A}_{p} $ or $ | 0 \rangle + | \zeta \rangle  $ which can be identified as vacuum, creates and annihilates the substantial photons through the interference.

Loosely speaking, the unobservable potentials are oriented by the polarizers such as Eq. (\ref{eq:90and0do}) or Eq. (\ref{eq:a2-phi}). Then the substantial photons surf on the sea of the oriented potentials which can change into substantial photons through the interference.

\section{New explanation for delayed choice quantum eraser}

\begin{figure}[t]
\centering
\includegraphics[width=18pc]{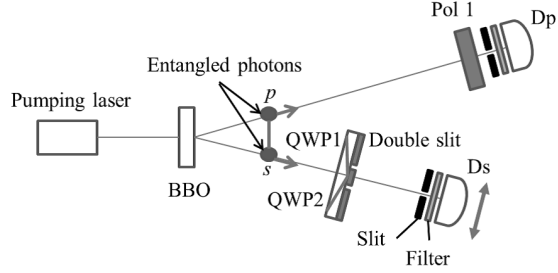}
\caption{Typical setup for the Delayed Choice Quantum Eraser. QWP1 and QWP2 are quarter-wave plates aligned in front of the double slit with fast axes perpendicular. Pol1 is a linear polarizer. BBO ($\beta {\rm -BaB_{2}O_{4}} $) crystal generates entangled photons by spontaneous parametric down-conversion \cite{PhysRevA.65.033818}.}
\label{fig:DCQE} 
\end{figure}

In this section, we show new explanation for Delayed Choice Quantum Eraser as shown in Figure \ref{fig:DCQE} which consists of an entangled photon source and two detectors. The delayed choice has been demonstrated when the distance from BBO to polarizer 1 is longer than that from BBO to the double slit \cite{PhysRevA.65.033818}.

Here we should take particular note of the fact that the polarization angle of polarizer 1 has been chosen before the entangled photons are generated. Walborn et al. \cite{PhysRevA.65.033818} have pointed out that ``the experiment did not allow for the observer to choose the polarization angle in the time period after photon {\it s} was detected and before detection of {\it p}''. From the principle of causality, their point will be reasonable.

However, mathematical description for the phenomenon requires entangled state such as
\begin{equation}
| \psi \rangle = \frac{1}{\sqrt{2}} \left( | x \rangle_{\it s} | y \rangle_{\it p} + | y \rangle_{\it s} | x \rangle_{\it p} \right)
\label{eq:entangle}
\end{equation}
The entangled state declares that the state of the whole system is a quantum-superposition state consisting of $ | x \rangle_{\it s} | y \rangle_{\it p} $ and $ | y \rangle_{\it s} | x \rangle_{\it p} $.
Therefore, when the state of one photon ({\it s} or {\it p}) is observed and determined to be $ | x \rangle$, that of the other photon ({\it p} or {\it s}) suddenly changes from the quantum-superposition state into $ | y \rangle $ even if the photons separate from each other, which postulates the existence of long-range correlation beyond the causality (spooky action at a distance).
This postulate represents a critical defect and serious problem with the traditional explanations as pointed out in a paper by ``Einstein, Podolsky and Rosen (EPR)'' \cite{EPR}.

Hence we grapple with a strange physical phenomenon from the moment that we choose the polarization angle of polarizer 1 to the moment BBO generates the entangle photon pairs.

The unobservable potentials, which can change from the potentials into substantial photons, eternally populate the whole space not forgetting the space between BBO and Polarizer 1 independent of substantial photons. Hence, the space will be populated by the unobservable potentials which are oriented by polarizer 1 as described above. More precisely, the potentials determine the polarization of substantial photons in the space in advance depending on the polarization angle of polarizer 1.

For example, if we choose the polarization angle of polarizer 1 to $ \phi $ which is measured from the polarization angle $ \psi $ of created photons, then the unobservable potential is oriented to $  |0 \rangle + | \zeta_{\phi} \rangle = |0 \rangle + \frac{1}{2} \gamma e^ {i  (\phi - \psi)} e^{i \theta/2}  | 0 \rangle - \frac{1}{2} \gamma e^{- i  (\phi - \psi)} e^{- i \theta/2}  | 0 \rangle $ at polarizer 1 and propagates to BBO. BBO is forced to generate the photon pair with polarization {\it p} : $ \phi $ and {\it s} : $ \phi \pm \frac{1}{2} \pi $ according to the arrival potentials. The mathematical description is as follows. By applying a photon creation operator $ \hat{A_{\psi}}^{\dagger} $ to the polarized potentials, i. e.,
\begin{equation}
\hat{A_{\psi}}^{\dagger} |0 \rangle + \hat{A_{\psi}}^{\dagger} | \zeta_{\phi} \rangle = | \psi \rangle + \frac{1}{2} \gamma e^ {i (\phi - \psi) } e^{i \theta/2}  | \psi \rangle - \frac{1}{2} \gamma e^{- i (\phi - \psi)} e^{- i \theta/2}  | \psi \rangle
\label{eq:polpot} 
\end{equation}
Equation (\ref{eq:polpot}) can be calculated as the created photon state at BBO. Then the intensity of the created photon can be calculated in this setup ($ \theta = 0 $) as follows.
\begin{equation}
\langle I \rangle \propto \frac{1}{2} + \frac{1}{2} \cos \left( 2 \phi - 2 \psi \right)
\end{equation}
In order to create a photon, i. e., $ \langle I \rangle = 1 $, $ \psi = \phi $ will be required.

Then the polarization of the photon pair is fixed by the unobservable potentials instead of the entangle state in Eq. (\ref{eq:entangle}). Therefore, when the polarization angle is set to the fast axis of QWP (Quarter-wave plate) 1 or 2, the interference pattern can be observed.

In this case, we are not aware of the determination of the polarization of the photon pair by the unobservable potentials. This is the reason why the state seems to be ``entangled'', and the choice of the polarization angle of polarizer 1 seems to be ``delayed''.

In order to confirm the new explanation, we should make experiments with a shutter between BBO and polarizer 1 as follows. First, close the shutter not to make a definite orientation of the unobservable potentials. After the entangled photon pairs are generated, open the shutter. When the photon {\it s} is detected by Ds, close the shutter again. After a time period, we excite BBO to generate the next entangled photon pairs. When the next pairs are generated, open the shutter again. By repeating these procedures, we can make a comparison between the traditional results and new result. If the definite orientation of the unobservable potentials as mentioned above is valid, no interference pattern can be observed even if the polarization angle of Polarizer 1 is set to the fast axis of QWP 1 or 2 throughout the experiment.

Note that because the unobservable potentials obey Maxwell equations propagate at the speed of light, the above time period that prevents the unobservable potentials from being oriented should be longer than the distance between BBO and the shutter divided by the speed of light.

The above new explanation is based on the preselected polarization by the setup. However, even if the polarizations of the photon pair are randomly selected, the measurement results seem to have the long-range correlation beyond the causality as follows. From Eq. (\ref{eq:90and0do}), the measurement results of photons $ {\it s} $ and $ {\it p} $ are expressed as follows.
\begin{eqnarray}
\langle I_{\it s} \rangle \propto \frac{1}{2} + \frac{1}{2} \cos\left( 2 \phi \right) & , \ \ \ & \langle I_{\it p} \rangle \propto \frac{1}{2} - \frac{1}{2} \cos\left( 2 \phi \right)
\label{eq:intensity-sp}
\end{eqnarray}
There is no such a classical correlation. The above results are identical to the traditional quantum-mechanical predictions and violate Bell's inequalities.
Therefore, the long-range correlation associated with the interference between the photons and unobservable potentials is observed in all the experimental setups presented here. This is the answer to the so called ``setting-independence loophole'' \cite{CosmicBellTest}.

Therefore, the confirmation method for the preselected polarization case described above has to be carefully implemented. When there are no polarizers, the polarization is randomly selected. Hence, a detection frequency of photons by $ {\rm D_{p}} $ proportional to the intensity of measured photon will be extremely lower than the case when there are polarizers. The difference of the detection frequency will be the only way to distinguish the new explanation from traditional one.

\section{Tensor form of the electromagnetic fields}
We have introduced the operator by using $ \gamma^{2} = -1 $ such as Eq. (\ref{eq:a2-phi}), which expresses the unobservable potentials for convenience in calculation in the above. When we use tensor form of the electromagnetic fields, the operator and results can be spontaneously obtained in following manner.
The followings is almost the same as the description for the single photon interference in Ref.\cite{Morimoto}. 

The electromagnetic potentials are expressed as following four-vector in Minkowski space.
\begin{equation}
A^{\mu} = (A^{0}, \ A^{1}, \ A^{2}, \ A^{3} ) = (\phi /c, \ {\bf A})
\end{equation}
The  four-current is also expressed as following four-vector.
\begin{equation}
j^{\mu} = (j^{0}, \ j^{1}, \ j^{2}, \ j^{3}) = (c \rho, \ {\bf i})
\end{equation}
When we set the axises of Minkowski space to $ x^{0} = ct $, $ x^{1} = x, \ x^{2} = y, \ x^{3}  =z $, Maxwell equations with Lorentz condition are expressed as follows.
\begin{eqnarray}
\Box A^{\mu} = \mu_{0} j^{\mu} &, \ \ \ & \partial_{\mu} A^{\mu} = 0
\label{eq:Max-Lorentz}
\end{eqnarray}
In addition, the conservation of charge $ {\rm div} \ {\bf i} + \partial \rho / \partial t = 0 $ is expressed as $ \partial_{\mu} j^{\mu} = 0 $, where $ \partial_{\mu} = ( 1/c\partial t , \ 1/\partial x , \ 1/\partial y , \ 1/\partial z ) = ( 1/\partial x^{0} , \ 1/\partial x^{1} , \ 1/\partial x^{2} , \ 1/\partial x^{3} ) $, and  $ \Box $ stands for the d'Alembertian: $ \Box \equiv \partial_{\mu} \partial^{\mu} \equiv \partial^2 / c^2 \partial t^2 - \Delta $.

The transformation between covariance and contravariance vector can be calculated by using the simplest form of Minkowski metric tensor $ \texttt{g}_{\mu \nu} $ as follows.
\begin{eqnarray}
 \texttt{g}_{\mu \nu} = \texttt{g}^{\mu \nu} & = & \left[ 
 \begin{array}{cccc}
 1 & 0 & 0 & 0 \\
0 & -1 & 0 & 0 \\
0 & 0 & -1 & 0 \\
0 & 0 & 0 & -1\\
 \end{array} 
\right] \nonumber \\
A_{\mu} = \texttt{g}_{\mu \nu} A^{\nu} &, \ \ \ & A^{\mu} = \texttt{g}^{\mu \nu} A_{\nu}
\end{eqnarray} 
The following quadratic form of four-vectors is invariant under a Lorentz transformation.
\begin{equation}
(x^{0} )^{2} - ( x^{1} )^{2} - (x^{2} )^{2} - (x^{3} )^{2} 
\end{equation}
The above quadratic form applied a minus sign expresses the wave front equation and can be described by using metric tensor.
\begin{equation}
- \texttt{g}_{\mu \nu} x^{\mu} x^{\nu} = - x^{\mu} x_{\mu} = x^{2} + y^{2} + z^{2} - c^{2} t^{2} = 0
\end{equation}
This quadratic form which includes minus sign is also introduced to inner product of arbitrarily vectors and commutation relations in Minkowski space.

The four-vector potential satisfied Maxwell equations with vanishing the four-vector current can be expressed as following Fourier transform in terms of plane wave solutions \cite{QFT}.
\begin{equation}
A_{\mu} (x) = \int d \tilde{k} \sum^{3}_{\lambda = 0} [a^{(\lambda)} (k) \epsilon^{(\lambda)}_{\mu} (k) e^{- i k \cdot x } + a^{(\lambda) \dagger} (k) \epsilon^{(\lambda) *}_{\mu} (k) e^{ i k \cdot x} ]
\end{equation}
\begin{equation}
\tilde{k} = \frac{d^{3} k}{2 k_{0} (2 \pi )^{3}} \ \ \ k_{0} = | {\bf k} | 
\end{equation}
where the unit vector of time-axis direction $ n $ and polarization vectors $ \epsilon^{(\lambda )}_{\mu} (k) $ are introduced as $ n^{2} = 1, \ n^{0}>0 $ and $ \epsilon^{(0)} = n $, $ \epsilon^{(1)} $ and $ \epsilon^{(2)} $ are in the plane orthogonal to $k$ and $n$
\begin{equation}
\epsilon^{(\lambda)} (k) \cdot \epsilon^{(\lambda ')} (k) = - \delta_{\lambda , \lambda '} \ \ \ \  \lambda \ , \  \lambda ' = 1, \ 2
\end{equation}
$ \epsilon^{(3)} $ is in the plane $ (k, \ n) $ orthogonal to $n$ and normalized
\begin{equation}
\epsilon^{(3)} (k) \cdot n = 0, \ \ \ \ [ \epsilon^{(3)} (k) ]^2 = -1
\end{equation}

Then $ \epsilon^{(0)} $ can be recognized as a polarization vector of scalar waves, $ \epsilon^{(1)} $ and $ \epsilon^{(2)} $ of transversal waves and $ \epsilon^{(3)} $ of a longitudinal wave. Then we take these vectors as following the easiest forms.
\begin{equation}
\epsilon^{(0)} = \left(
\begin{array}{l}
1 \\
0 \\
0 \\
0
\end{array}
\right)
\ \ \ 
\epsilon^{(1)} = \left(
\begin{array}{l}
0 \\
1 \\
0 \\
0
\end{array}
\right)
\ \ \ 
\epsilon^{(2)} = \left(
\begin{array}{l}
0 \\
0 \\
1 \\
0
\end{array}
\right)
\ \ \ 
\epsilon^{(3)} = \left(
\begin{array}{l}
0 \\
0 \\
0 \\
1
\end{array}
\right)
\end{equation}

When the Fourier coefficients of the four-vector potentials are replaced by operators as $ \hat{A}_{\mu} \equiv \sum^{3}_{\lambda = 0} \hat{a}^{(\lambda)} (k) \epsilon^{(\lambda)}_{\mu} (k) $, the commutation relations are obtained as follows.
\begin{equation}
[ \hat{A}_{\mu} (k), \ \hat{A}_{\nu}^{\dagger} (k') ] = - \texttt{g}_{\mu \nu} \delta (k-k')
\end{equation}
The time-axis component (corresponds to $ \mu , \nu = 0 $ scalar wave, i. e., scalar potential because $ \epsilon^{(0)}_{\mu} (k) = 0 \ (\mu \neq 0)  $) has the opposite sign of the space axes. Because $ \langle 0 | \hat{A}_{0} (k)  \hat{A}_{0}^{\dagger} (k') | 0 \rangle = - \delta (k-k') $,
\begin{equation}
\langle 1 | 1 \rangle = - \langle 0 | 0 \rangle \int d \tilde{k} | f(k) |^{2}
\end{equation}
where $ | 1 \rangle = \int d \tilde{k} f(k) \hat{A}_{0}^{\dagger} (k) | 0 \rangle $. Therefore, the time-axis component is the root cause of indefinite metric. Note that the products of the operators replaced from the four-vectors must introduce the same formalism.
\begin{equation}
\hat{A}^{\dagger} \hat{A} = - \texttt{g}_{\mu \nu} \hat{A}^{\mu \dagger} \hat{A}^{\nu} 
\end{equation}
In order to utilize the indefinite metric as follows, Coulomb gauge that removes the scalar potentials should not be used.

Here we can recognize the potentials before passing through the polarizers 1 and 2 as 
\begin{equation}
A_{\mu} = (A_{0}, \ A_{1}, \ A_{2}, \ 0)
\end{equation}
where, we neglect the longitudinal wave which is considered to be unphysical presence, i. e., $ A_{3} = 0 $ for simplicity.
When there are an x-polarized photon and scalar potential which pass through each polarizer, the potentials passing through the polarizers can be expressed as
\begin{eqnarray}
A_{({\rm x \ pol \ 1}) \ \mu} = \left(\frac{1}{2} e^{i \theta_{x}/2}A_{(x)0}, \ A_{(x)1}, \ 0, \ 0 \right) & , \ \ \ & A_{({\rm x \ pol \ 2}) \ \mu} = \left(\frac{1}{2} e^{-i \theta_{x}/2}A_{(x)0}, \ 0, \ 0, \ 0 \right) 
\label{eq:Axpol}
\end{eqnarray}
When these scalar potentials undergo a $ | \phi | $ phase shift, i. e., the angle of the polarizer 3, by passing through the polarizer 3, the phase terms will be shifted to $\pm i \left( | \phi |  + \theta_{x}/2 \right) $. 
 Here we identify the number operators as $ \langle 1 | A_{0}^{\dagger} A_{0} | 1 \rangle = \langle 1 | A_{1}^{\dagger} A_{1} | 1 \rangle = \langle 1 | A_{2}^{\dagger} A_{2} | 1 \rangle = 1 $ because of the Lorentz invariance.
Hence the single photon interference in Eq. (\ref{eq:intensity-x}) or (\ref{eq:intensity-sp}) is obtained as followings.
\begin{equation}
A_{({\rm x \ pol \ 1, \ 2 \rightarrow 3}) \ \mu} \equiv A_{({\rm x \ pol \ 1\rightarrow 3}) \ \mu} + A_{({\rm x \ pol \ 2\rightarrow 3}) \ \mu} = \left(\cos (| \phi | + \frac{\theta_{x}}{2} ) A_{(x)0}, \ A_{(x)1}, \ 0, \ 0 \right) 
\end{equation}
\begin{equation}
\langle I_{\it s} \rangle \propto \langle 1 | A_{({\rm x \ pol \ 1, \ 2 \rightarrow 3})}^{\dagger}  A_{({\rm x \ pol \ 1, \ 2 \rightarrow 3})} | 1 \rangle = \frac{1}{2} - \frac{1}{2} \cos (2 | \phi | + \theta_{x})
\end{equation}
Similarly, in the case of a y-polarized photon
\begin{eqnarray}
A_{({\rm y \ pol \ 1}) \ \mu} = \left( \frac{1}{2} e^{i \theta_{y}/2}A_{(y)0}, \ 0, \ 0, \ 0 \right) &, \ \ \ & A_{({\rm y \ pol \ 2}) \ \mu} = \left( \frac{1}{2} e^{-i \theta_{y}/2}A_{(y)0}, \ 0, \ A_{(y)2}, \ 0 \right)
\label{eq:Aypol}
\end{eqnarray}
\begin{equation}
A_{({\rm y \ pol \ 1, \ 2 \rightarrow 3}) \ \mu} \equiv A_{({\rm y \ pol \ 1\rightarrow 3}) \ \mu} + A_{({\rm y \ pol \ 2\rightarrow 3}) \ \mu} = \left( \cos (| \phi | + \frac{\theta_{y}}{2} ) A_{(y)0}, \ 0, \ A_{(y)2}, \ 0 \right)
\end{equation}
Then
\begin{equation}
\langle I_{\it p} \rangle \propto \langle 1 | A_{({\rm y \ pol \ 1, \ 2 \rightarrow 3})}^{\dagger}  A_{({\rm y \ pol \ 1, \ 2 \rightarrow 3})} | 1 \rangle = \frac{1}{2} - \frac{1}{2} \cos (2 | \phi | + \theta_{y})
\end{equation}
By choosing $ \theta \equiv \theta_{x} = - ( \theta_{y} + \pi ) $, i. e., the potentials undergo $ \pi $ phase shift and the relatively-same phase shift at polarizers 1 and 2 when divided,
\begin{eqnarray}
\langle I_{\it s} \rangle \propto \frac{1}{2} - \frac{1}{2} \cos (2 | \phi | + \theta) &, \ \ \ & \langle I_{\it p} \rangle \propto \frac{1}{2} + \frac{1}{2} \cos (2 | \phi | - \theta)
\end{eqnarray}
Hence, we should choose $ \theta = \theta + \pi $ to correct the reversed signs, which is attributed to the difference between using $ \gamma^{2} = -1 $ and tensor form.

In case of both polarization photons exist, the potentials just before polarizer 3 will be expressed by summation of Eqs. (\ref{eq:Axpol}) and (\ref{eq:Aypol}). Then the potentials that undergo a $ | \phi | $ phase shift by polarizer 3 can be expressed as follows.
\begin{equation}
A_{({\rm x, \ y \ pol \ 1, \ 2 \rightarrow 3}) \ \mu}  = \left( A_{(x)0} \cos ( | \phi | + \frac{\theta_{x}}{2}) + A_{(y)0} \cos (| \phi | + \frac{\theta_{y}}{2} ) , \ A_{(x)1}, \ A_{(y)2}, \ 0 \right)
\end{equation}

Therefore, the photon number operator of the output of the polarizer 3 can be calculated as follows.
\begin{eqnarray}
& & A_{({ \rm x, \ y \ pol \ 1, \ 2 \rightarrow 3})}^{\dagger} A_{({ \rm x, \ y \ pol \ 1, \ 2 \rightarrow 3})} \nonumber \\
& = &- A_{(x)0}^{\dagger} A_{(x)0} \cos^{2} ( | \phi | + \frac{\theta_{x}}{2} ) - A_{(y)0}^{\dagger} A_{(y)0} \cos^{2} ( | \phi | + \frac{\theta_{y}}{2} ) ) + A_{(x)1}^{\dagger} A_{(x)1} + A_{(y)2}^{\dagger} A_{(y)2} \nonumber \\
& & - ( A_{(x)0}^{\dagger} A_{(y)0} + A_{(y)0}^{\dagger} A_{(x)0} ) \cos ( | \phi | + \frac{\theta_{x}}{2} ) \cos ( | \phi | + \frac{\theta_{y}}{2} ) 
\end{eqnarray}

Then by choosing $ \theta \equiv \theta_{x} = - ( \theta_{y} + \pi ) $,
\begin{eqnarray}
& & \langle 1 | A_{({ \rm x, \ y \ pol \ 1, \ 2 \rightarrow 3})}^{\dagger} A_{({ \rm x, \ y \ pol \ 1, \ 2 \rightarrow 3})} | 1 \rangle \nonumber \\
& = &1 - \frac{1}{2} \cos (2 | \phi | + \theta ) + \frac{1}{2} \cos (2 | \phi | - \theta ) \nonumber \\
& & -  \langle 1 | ( A_{(x)0}^{\dagger} A_{(y)0} + A_{(y)0}^{\dagger} A_{(x)0} ) | 1 \rangle \cos ( | \phi | + \frac{\theta}{2} ) \sin ( | \phi | - \frac{\theta}{2} )
\label{eq:interfare-both-final}
\end{eqnarray}
Here we should recognize $ | 1 \rangle = ( | 1 \rangle_{x} + | 1 \rangle_{y} ) $ as mentioned above, and $ A_{(x)0} $ and $ A_{(y)0} $ annihilate x and y-polarized photon, respectively, i. e., $ A_{(x)0} | 1 \rangle = | 0 \rangle_{x} $ and $ A_{(y)0} | 1 \rangle = | 0 \rangle_{y} $. Because $ _{x} \langle 0 | 0 \rangle_{y} = 0 $,
\begin{equation}
-  \langle 1 | ( A_{(x)0}^{\dagger} A_{(y)0} + A_{(y)0}^{\dagger} A_{(x)0} ) | 1 \rangle =  0
\end{equation}

Hence Eq. (\ref{eq:interfare-both-final}) corresponds to Eqs. (\ref{eq:intensity-xy-comp}) and (\ref{eq:intensity-xy-Heisen}) except the $ \pi $ phase shift of $ \theta $.

\section{Discussion}
In this paper, we have taken advantage of the indefinite metric property of scalar potentials. Here we discuss what the scalar field represents.

Usually in quantum optics, we can split the electric field  and current density by using Coulomb gauge as follows \cite{Loudon}. 
\begin{eqnarray}
{\bf E} = {\bf E}_{\rm T} + {\bf E}_{\rm L}, \ \ \ & \nabla \cdot {\bf E}_{ \rm T} = 0, \ \ \ & \nabla \times {\bf E}_{ \rm L} = 0 \nonumber \\
{\bf i} = {\bf i}_{\rm T} + {\bf i}_{\rm L}, \ \ \ & \nabla \cdot {\bf i}_{ \rm T} = 0, \ \ \ & \nabla \times {\bf i}_{ \rm L} = 0
\label{eq:trans-and long}
\end{eqnarray}
where the indexes ``T'' and ``L'' stand for ``Transverse'' and ``Longitudinal'', respectively. By using electromagnetic potentials, ``Transverse''components of Maxwell equations can be described as follows.
\begin{eqnarray}
\nabla \times {\bf E}_{ \rm T} = - \frac{\partial {\bf B}}{\partial t} & , \ \ \ & \nabla \times {\bf B} = \frac{1}{c^{2}} \frac{\partial {\bf E}_{\rm T}}{\partial t} + \mu_{0} {\bf i}_{\rm T} \nonumber \\
{\bf E}_{ \rm T}  =  -\frac{\partial {\bf A}}{\partial t} & , \ \ \ & \nabla \cdot {\bf B} = 0
\label{eq:maxwell-coulomb-trans}
\end{eqnarray}
where $ {\bf B} $ is the magnetic field. We can also obtain following ``Longitudinal'' components.
\begin{eqnarray}
{\bf E}_{ \rm L}  =  -\nabla \phi &, \ \ \ & \nabla \cdot {\bf E}_{ \rm L} = \frac{\rho}{\epsilon_{0}} \nonumber \\
{\bf i}_{ \rm L}  =  \epsilon_{0} \nabla \frac{\partial \phi}{\partial t} & = & - \epsilon_{0} \frac{\partial {\bf E}_{\rm L}}{\partial t}
\label{eq:maxwell-coulomb-long}
\end{eqnarray}
Hence the transverse component seems associated with the magnetic field variation, and the longitudinal component seems associated with charges as the regular scalar potential.

However, these associations are justified in a particular coordinate system, i. e., ``relative associations''. When the coordinate system is changed according to Lorentz transformation, ``Transverse'' and ``Longitudinal'' components are mixed. Then the associations have no meaning which is the important assertion of relativity \cite{Einstein}. This is why we equate scalar potentials with vector potentials, i. e., identify the number operators as $ \langle 1 | A_{0}^{\dagger} A_{0} | 1 \rangle = \langle 1 | A_{1}^{\dagger} A_{1} | 1 \rangle = \langle 1 | A_{2}^{\dagger} A_{2} | 1 \rangle = 1 $ by Lorentz invariance. In addition, the Coulomb gauge removes the explicit covariance of Maxwell equations. Hence we would better use Maxwell equations (\ref{eq:Max-Lorentz}) with Lorentz gauge.
By utilizing the linearity of  the equation (\ref{eq:Max-Lorentz}), we can express Maxwell equations with Lorentz condition as follows.
\begin{equation}
\begin{array}{lclcl}
\Box A^{\mu} & = & \Box ( A_{\rm (mat)}^{\mu} + A_{\rm (vac)}^{\mu} )  & = &  \mu_{0} j^{\mu} \\
\partial_{\mu} A^{\mu} & = & \partial_{\mu}  ( A_{\rm (mat)}^{\mu} + A_{\rm (vac)}^{\mu} ) & = & 0
\label{eq:Max-vac+mat}
\end{array}
\end{equation}
where index ``mat'' and ``vac'' mean ``matter'' associated with four-current and ``vacuum'', respectively. If we naturally assume that there are no four-current in vacuum, then $ A_{\rm (mat)}^{\mu} $ and $ A_{\rm (vac)}^{\mu} $ obey the following Maxwell equations respectively.
\begin{eqnarray}
\Box A_{\rm (mat)}^{\mu} = \mu_{0} j^{\mu} &, \ \ \ & \partial_{\mu} A_{\rm (mat)}^{\mu} = 0
\label{eq:Max-mat}
\end{eqnarray}
\begin{eqnarray}
\Box A_{\rm (vac)}^{\mu} = 0 &, \ \ \ & \partial_{\mu}  A_{\rm (vac)}^{\mu} = 0
\label{eq:Max-vac}
\end{eqnarray}

Equation (\ref{eq:Max-mat}) will express substantial photon excited by the four-current. Note that when we consider the spatial domain far from and exclude the four-current, Equation (\ref{eq:Max-vac}) replacing $ A_{\rm (vac)}^{\mu} $ with $ A_{\rm (mat)}^{\mu} $ can express the motion of the potentials in the domain associated with the four-current. 

In contrast, Equation (\ref{eq:Max-vac}) expresses the motion of the potentials unrelated to  ``matter'' in vacuum. Therefore, we can imagine that vacuum is the sea filled with unobservable potentials, which evokes the concept of an ether. Although the static ether has been rejected by special relativity \cite{Einstein}, the above filling potentials are not static entity but propagate at the speed of light. Aharonov-Bohm effect clearly presents that the unobservable potentials without electromagnetic field can cause electron interference \cite{ABeffe,Tonomura1,Tonomura2}. By the same token, the filling potentials in Eq. (\ref{eq:Max-vac}) can cause interference with substantial photon, Eq. (\ref{eq:Max-mat}) as if it were a local oscillator for homodyne detection attached to space-time as discussed in Ref.\cite{Morimoto}.

We generally calculate photon related phenomena using $ A^{\mu} $ in Eq. (\ref{eq:Max-vac+mat}) unconsciously, i. e., without separation into ``matter'' and ``vacuum''. However, we can not distinguish $ A_{\rm (mat)}^{\mu} $ from $ A_{\rm (vac)}^{\mu} $, which is very much like distinguish sea spray from seawater. Indeed, no separation will be required because both are ever-changing potentials derived from the same Maxwell equation (\ref{eq:Max-vac+mat}).
Therefore, the filling potentials in vacuum can expel and incorporate the potentials associated with ``matter'', which makes us imagine that vacuum can create and annihilate substantial photon.

The scalar field used in this paper correspond to the scalar component of this filling potentials in the literature.

Although we estimate that the origin of the filling potentials might be the fabric of space-time from the above consideration, the investigation will be a subject for a further study.

\section{Conclusions}
We have presented that the quantum eraser can be explained without quantum-superposition states by introducing the unobservable (scalar) potentials whose probability (or more like ``interference'') amplitudes are zero. The explanation presents the concept that vacuum can create and annihilate the substantial photons.

We have also investigated the delayed choice experiment under the assumption that the polarization of the photon pairs is determined by the unobservable (scalar) potentials which are oriented by the setup of the experiment in advance. Moreover, we show that the interference between the photons and unobservable potentials makes the long-range correlation beyond the causality that does not really exist in nature but seems to exist regardless of the assumption. In addition to these discussions based on a method for convenience in calculation, we have shown rigorous mathematical treatment using tensor form (covariant quantization).

The new explanations obtained in the present paper are more general and physically consistent than traditional explanations which require paradoxical quantum-superposition states and entangled states.
The other experiments and considerations have been reported, which seem like paradoxes \cite{PhysRevLett.100.220404,PhysRevLett.84.1,aspect1981experimental,aspect1982experimental,aspect1982experimental-2,clauser1974experimental,CosmicBellTest}. We believe that the paradoxes can be avoided by the new explanation. Moreover, we should investigate whether engineering applications based on wave packet reduction or entangled states are feasible technologies or not, because an inevitable conclusion by the rigorous derivation described in this paper can remove the paradoxical base concepts of the applications.

The new explanation presented here and Ref.\cite{Morimoto} compel a restructuring of the traditional standard quantum theory. However, this is the real natural law without the enigmatic and paradoxical thought processes such as quantum-superposition and entanglement based on the ``probabilistic interpretation''.

\section*{Acknowledgment}
The author would like to thank K. Sato and Dr. S. Takasaka for their helpful discussions.



%

\end{document}